\documentclass[preprint,showpacs,nofootinbib]{revtex4-1}

\usepackage{graphicx}
\usepackage{bm}
\usepackage[english]{babel}
\usepackage{dcolumn}
\usepackage{amssymb}

\begin{document}

\title{Improvement of the envelope theory with the dominantly orbital state method}

\author{Claude Semay}

\email[E-mail: ]{claude.semay@umons.ac.be}
\affiliation{Service de Physique Nucl\'{e}aire et Subnucl\'{e}aire,
Universit\'{e} de Mons,
UMONS Research Institute for Complex Systems,
Place du Parc 20, 7000 Mons, Belgium}
\date{\today}

\date{\today}

\begin{abstract}
The envelope theory, also known as the auxiliary field method, is a simple technique to compute approximate solutions of Hamiltonians for $N$ identical particles in $D$ dimensions. The quality of the approximate eigenvalues can be improved by adding a free parameter in the characteristic global quantum number of the solutions. A method is proposed to determine the value of this parameter by comparing the eigenvalues computed with the envelope theory to the corresponding ones computed with a $N$-body generalization of the dominantly orbital state method. The accuracy of the procedure is tested with several systems. 
\end{abstract}
%\keywords{Bound states \and Many-body systems \and Envelope theory \and Dominantly orbital state method}
%\PACS{03.65.Ge}
%03.65.Ge Solutions of wave equations: bound states

\maketitle

\section{Introduction}
\label{sec:intro}

The envelope theory (ET), also known as the auxiliary field method, is a procedure to obtain approximate solutions of $N$-body Hamiltonians in quantum mechanics \cite{hall80,hall04,silv10,sema13a}. In the most favorable cases, an approximate eigenvalue is an analytical lower or upper bound. The method relies on the replacement of the Hamiltonian $H$ under study by an auxiliary Hamiltonian $\widetilde H$ which is solvable, the eigenvalues of $\widetilde H$ being optimized to be as close as possible to those of $H$. Recently, the accuracy of the ET for eigenvalues and eigenvectors has been tested by computing the ground state of various systems containing up to 8 bosons \cite{sema15}. This comparison was possible thanks to accurate numerical results published in \cite{horn14} and obtained by using explicitly correlated Gaussian basis \cite{suzu98}.

In several cases, it has been shown that the approximate eigenvalues computed with the ET can be improved by modifying the structure of the characteristic global quantum number of the method \cite{silv10,silv12}. The price to pay is generally the loss of the variational character of the approximation. In \cite{sema15}, a modification of the global quantum number is proposed and tested by introducing only one supplementary parameter $\phi$. This procedure relies on the result obtained in \cite{loba09}, where a universal effective quantum number for centrally symmetric 2-body systems is proposed. The new parameter $\phi$ can be determined by a fit on analytical results for $N=2$ or on accurate numerical results \cite{sema15}. 

In this paper, we propose a method to compute the value of $\phi$, independently of the knowledge of analytical or numerical previous calculations. The idea is to compare the eigenvalues computed with the ET to the corresponding ones computed with a $N$-body generalization of the dominantly orbital state (DOS) method. In this last method, an  approximate solutions is found by quantizing the radial motion around a semiclassical solution for a circular motion. Developed at the origin for 2-body systems \cite{olss97}, it has been extended to 3-body systems \cite{sema13b}.

The paper is organized as follows. The ET method is briefly recalled in Sec.~\ref{sec:et}. The extension of the DOS method for $N$-body systems is presented in Sec.~\ref{sec:dos}. In Sec.~\ref{sec:phi}, the computation of the parameter $\phi$ is described. The relevance of this parameter is tested with various systems in Sec.~\ref{sec:res}. Since the behavior of observables is not really predictable when $\phi$ is varied \cite{sema15}, we do not consider eigenvectors here. Some concluding remarks are finally given. 

\section{The envelope theory}
\label{sec:et}

We consider the following general Hamiltonian, in a $D$ dimensional space ($D \ge 2$), for $N$ identical particles 
\begin{equation}
\label{HNb}
H=\sum_{i=1}^N T(|\bm p_i|) + \sum_{i=1}^N U\left(|\bm r_i - \bm R|\right) + \sum_{i\le j=1}^N V\left(|\bm r_i - \bm r_j|\right).
\end{equation}
$T$ is a kinetic energy, $U$ a one-body interaction and $V$ a two-body  potential ($\hbar=c=1$). The center of mass motion is removed ($\sum_{i=1}^N \bm p_i = \bm 0$) and $\bm R = \frac{1}{N}\sum_{i=1}^N \bm r_i$ is the center of mass position. It is shown in \cite{sema13a} that, in the framework of the ET, an approximate eigenvalue $E$ is given by the following set of equations for a completely (anti)symmetrized state:
\begin{eqnarray}
\label{AFM1N}
&&E=N\, T(p_0) + N\, U \left( \frac{r_0}{N} \right) + C_N\, V \left( \frac{r_0}{\sqrt{C_N}} \right), \\
\label{AFM2N}
&&r_0\, p_0=Q, \\
\label{AFM3N}
&&N\, p_0\, T'(p_0) =  r_0\, U' \left( \frac{r_0}{N} \right) + \sqrt{C_N}\, r_0\, V' \left( \frac{r_0}{\sqrt{C_N}} \right),
\end{eqnarray}
where $W'(x)=dW/dx$, $C_N=N(N-1)/2$ is the number of particle pairs, and 
\begin{equation}
\label{QN}
Q = \sum_{i=1}^{N-1} (2 n_i + l_i) + (N - 1)\frac{D}{2}
\end{equation}
is a global quantum number. This corresponds to $N-1$ identical harmonic oscillators, implying a strong degeneracy. Following the forms of $T$, $U$ and $V$, the approximate value $E$ can have a variational character \cite{sema13a}. 

The method is easy to implement since it reduces to find the solution of a transcendental equation. Equations~(\ref{AFM1N}-\ref{AFM3N}) have a nice semiclassical interpretation \cite{sema13a}, but the ET yields true quantum results: approximate eigenvalues and eigenvectors. In this paper, we focus on the eigenvalues. Information about the corresponding eigenvectors can be found in \cite{silv12,sema15}. Let us simply mention the following results giving a clear interpretation of the quantities $p_0$ and $r_0$ \cite{silv12}:
\begin{eqnarray}
\label{p0}
p_0^2 &=& \frac{1}{N}\left\langle \sum_{i=1}^N \bm p_i^2 \right\rangle, \\
\label{r0}
r_0^2 &=& N \left\langle \sum_{i=1}^N (\bm r_i - \bm R)^2 \right\rangle =
\left\langle \sum_{i<j=1}^N (\bm r_i - \bm r_j)^2 \right\rangle. 
\end{eqnarray}

\section{The $N$-body dominantly orbital state method}
\label{sec:dos}

In the DOS method for 2-body systems, an approximate solution is found by quantizing the radial motion around a semiclassical solution for a circular motion \cite{olss97}. In \cite{sema13b}, this method has been extended to 3-body systems, but it is mentioned that a generalization to $N$-body seems difficult. Nevertheless, this appears finally possible starting from (\ref{AFM1N}-\ref{AFM3N}), since a semiclassical interpretation of the ET is established in terms of circular motions of $N$ particles in a fully symmetrical configuration \cite{sema13a}.

Let us consider a pure circular motion for the $N$ particles with the energy $E_0$. We can rewrite (\ref{AFM2N}) as $r_0\, p_0=\lambda$ \cite{sema13a}, where the total angular momentum $\lambda$ will be specified later. But, it is expected that $\lambda \gg 1$ in order to justify the semiclassical approximation. $E_0$ can be computed from the system~(\ref{AFM1N}-\ref{AFM3N}) with $Q$ replaced by $\lambda$. Now, let us determine a solution with energy $E=E_0+\Delta E$ for a small collective radial perturbation around this circular solution. Following a similar procedure as the one developed in \cite{sema13b}, we build from (\ref{AFM1N}) a Hamiltonian $\Delta H$ ruling this radial motion by the replacements,
\begin{eqnarray}
\label{p0r0rep1}
r_0 &\to& r_0+\Delta r, \\
\label{p0r0rep2}
p_0 = \frac{\lambda}{r_0} &\to& \sqrt{p_r^2+\frac{\lambda^2}{(r_0+\Delta r)^2}},
\end{eqnarray}
where $p_r$ and $\Delta r$ are conjugate variables. Assuming that $\langle \Delta r\rangle \ll r_0$ and $\langle p_r\rangle \ll p_0$, a power expansion of $\Delta H$, limited to the lowest non-vanishing orders, gives
\begin{equation}
\label{DeltaH}
\Delta H \approx \frac{1}{2\,\mu} p_r^2 + \frac{k}{2} \Delta r^2,
\end{equation}
with 
\begin{eqnarray}
\label{muk1}
\mu &=& \frac{\lambda}{N\,r_0\,T'(\lambda/r_0)}, \\
\label{muk1}
k &=& \frac{N\,\lambda}{r_0^4}\left( 2\,r_0\,T'(\lambda/r_0)+ \lambda\,T''(\lambda/r_0) \right)+ \frac{1}{N} U''(r_0/N) + V''(r_0/\sqrt{C_N}).
\end{eqnarray}
The terms in $\Delta r$ are cancelled by (\ref{AFM3N}). The perturbed energy $\Delta E$ is then simply given by 
\begin{equation}
\label{DeltaEA}
\Delta E = A\,\nu \quad \textrm{with} \quad A=\sqrt{\frac{k}{\mu}} \quad \textrm{and} \quad \nu=\frac{1}{2},\, \frac{3}{2},\, \ldots.
\end{equation}
It is expected that $\nu \ll \lambda$ in order that $\Delta E \ll E_0$. Explicitly,
\begin{eqnarray}
\label{A}
A^2 &=& \frac{2\, N^2}{r_0^2}\,T'^2\left(\frac{\lambda}{r_0}\right) 
+ \frac{N^2\,\lambda}{r_0^3}\,T'\left(\frac{\lambda}{r_0}\right)\,T''\left(\frac{\lambda}{r_0}\right) \nonumber \\
&&+ \frac{r_0}{\lambda}\,T'\left(\frac{\lambda}{r_0}\right)\,U''\left(\frac{r_0}{N}\right)
+ \frac{N\,r_0}{\lambda}\,T'\left(\frac{\lambda}{r_0}\right)\,V''\left(\frac{r_0}{\sqrt{C_N}}\right),
\end{eqnarray}
with $r_0$ being a solution of the system~(\ref{AFM1N}-\ref{AFM3N}) where $Q$ is replaced by $\lambda$. To produce relevant results, quantum numbers $\lambda$ and $\nu$ must be specified. This work is done for $N=3$ in \cite{sema13b}. The general case is discussed in the following section. Let us recall that the results given by this method are expected to be relevant for completely (anti)symmetrized states \cite{sema13b}, provided $\nu \ll \lambda$ and $\lambda \gg 1$.

\section{Improvement of the global quantum number}
\label{sec:phi}

In \cite{loba09}, it is shown that an effective quantum number $q$ that determines with high accuracy the level ordering of centrally symmetric 2-body systems has the following form (for practical purposes, our definition of $\phi$ differs from the one used in \cite{loba09})
\begin{equation}
\label{nqnum}
q=\phi \left( n+\frac{1}{2} \right) +l+\frac{D-2}{2},
\end{equation}
where the number $\phi$ depends on the system. This structure comes from the separation of angular and radial motions. The DOS method relies on a similar separation for $N=2$ and 3 \cite{olss97,sema13b}. For $N$-body systems, we can try 
\begin{eqnarray}
\label{lambdanu}
\nu &=& \sum_{i=1}^{N-1} n_i + (N - 1)\frac{1}{2}, \\
\lambda &=&\sum_{i=1}^{N-1}l_i + (N - 1)\frac{D-2}{2}.
\end{eqnarray}
This is the case in \cite{sema13b} with $N=3$, where the quantum numbers $\nu/\sqrt{3}$ and $\lambda/\sqrt{3}$ are used due to a particular scaling of the conjugate variables. 

It is shown in \cite{sema15}, that the energies obtained by the ET can be significantly improved by replacing the quantum number $Q$ in the system~(\ref{AFM1N}-\ref{AFM3N}) by the new quantum number 
\begin{equation}
\label{QNphi}
Q_\phi = \phi\, \nu + \lambda,
\end{equation}
which partly breaks the strong degeneracy of $Q$. This is an obvious generalization of (\ref{nqnum}), but for $N-1$ degrees of freedom. Such an extension from 2 to $N$ particles is expected to be relevant because of the existence, in the framework of the ET, of relations linking the energy of a $N$-body system to the energy of the corresponding 2-body system by a rescaling of the global quantum numbers \cite{silv11}. The genuine ET, with its possible variational solutions, is then recovered with $\phi=2$. For other values of $\phi$, the variational character of the solution cannot be guaranteed. 

If the parameter $\phi$ is assumed constant, it can be determined by a fit on a single known accurate solution. This procedure is used in \cite{sema15} for some specific systems. It should be preferable to have procedure allowing the computation of $\phi$ without the prior knowledge on some solutions of the system. This is possible by combining the information coming from the ET and the DOS method. 

Let us write $Q_\phi=\lambda(1+\epsilon)$ with $\epsilon=\phi\,\nu/\lambda$. If $\nu=\epsilon=0$, the solution $E_0$ of the system~(\ref{AFM1N}-\ref{AFM3N}) is the unperturbed solution of the DOS method. We can now search for a solution when $\epsilon \ll 1$, by setting
\begin{eqnarray}
\label{p0r0rep2}
r_0 &\to& r_0+\Delta r, \\
\label{p0r0rep2}
p_0 = \frac{\lambda}{r_0} &\to& \frac{\lambda(1+\epsilon)}{r_0+\Delta r}.
\end{eqnarray}
An expansion to the non-vanishing lowest orders of (\ref{AFM3N}) yields a relation between $\epsilon$ and $\Delta r$. Using this link in an expansion to the non-vanishing lowest orders of (\ref{AFM1N}) finally gives
\begin{equation}
\label{DeltaEB}
\Delta E = B\,\epsilon,
\end{equation}
with $B=B_n/B_d$ and
\begin{eqnarray}
\label{B2}
B_n &=& \frac{N\, \lambda}{r_0} \left[ 2\,T'\left(\frac{\lambda}{r_0}\right) 
+ \frac{\lambda}{r_0}\,T''\left(\frac{\lambda}{r_0}\right) \right] 
\left[ U'\left(\frac{r_0}{N}\right) + \sqrt{C_N}\, V'\left(\frac{r_0}{\sqrt{C_N}}\right) \right] \nonumber \\
&&+ \lambda\,T'\left(\frac{\lambda}{r_0}\right) 
\left[ U''\left(\frac{r_0}{N}\right) + N\, V''\left(\frac{r_0}{\sqrt{C_N}}\right) \right], \\
\label{B3}
B_d &=& \frac{N\, \lambda}{r_0^2} \,T'\left(\frac{\lambda}{r_0}\right) 
+ \frac{N\, \lambda^2}{r_0^3} \,T''\left(\frac{\lambda}{r_0}\right)
+ U'\left(\frac{r_0}{N}\right) + \frac{r_0}{N}\,U''\left(\frac{r_0}{N}\right) \nonumber \\
&& + \sqrt{C_N}\, V'\left(\frac{r_0}{\sqrt{C_N}}\right) + r_0\, V''\left(\frac{r_0}{\sqrt{C_N}}\right).
\end{eqnarray}
Equations (\ref{DeltaEA}) and (\ref{DeltaEB}) correspond to the same expansion $\nu \ll \lambda$. So we can write
\begin{equation}
\label{phiAB}
\phi = \frac{\lambda\,A}{B}.
\end{equation}
The complete expression for $\phi$ is quite complicated. But, the computation of $\phi$ is straightforward once $r_0$ is determined. In some cases, the final expression can be very simple, as shown with the examples below.

The procedure to compute a numerical solution is the following:
\begin{enumerate}
  \item Fix the global quantum numbers $\nu$ and $\lambda$;
  \item Compute the solution $r_0$ of the system~(\ref{AFM2N}-\ref{AFM3N}) with $Q$ replaced by $\lambda$;
  \item Compute $\phi$ with (\ref{phiAB});
  \item Compute $Q_\phi = \phi\, \nu + \lambda$;
  \item Compute the solution $E$ of the system~(\ref{AFM1N}-\ref{AFM3N}) with $Q$ replaced by $Q\phi$.
\end{enumerate}
If $r_0(Q)$ can be determined analytically, $r_0(\lambda)$ is used for step 2 and $r_0(Q_\phi)$ is used for step 5. 

With the replacement of $Q$ by $Q_\phi$ in the ET equations, we can hope that some relevant information about the relative influence of radial and orbital motions are now captured in the ET. Nevertheless, $\phi$ is determined in a particular limit: $\nu \ll \lambda$ and $\lambda \gg 1$. It is not clear that the definition of $Q_\phi$ can yield good results for other ranges of quantum numbers. This will be tested in the next section.

\section{Results}
\label{sec:res}

In this section, we consider various systems of $N$ identical particles with the same mass $m$, for which analytical ET solutions can be found. The ET formalism developed is valid for bosons or fermions with any values of $D$, but comparisons with numerical results will only be performed for boson-like particles in the $D=3$ space, as in \cite{sema15}.  

\subsection{Power-law 2-body potential for nonrelativistic particles}
\label{sec:pl2nr}

In the first example, nonrelativistic particles interact via a 2-body power-law potential
\begin{equation}\label{Hpl2}
T(p)=\frac{p^2}{2 m}, \ U(s)=0, \ V(r)=\textrm{Sgn}(b)\,a\,r^b, 
\end{equation}
with $a>0$ and $0\ne b>-2$. The ET gives the following upper (lower) bound for the energy $E$ if $b \le 2$ ($b > 2$)
%\label{Hpl2r0}
%r_0&=& \left( \frac{2^{(2-b)/2}\,N^{b/2}\,Q_\phi^2}{(N-1)^{(2-b)/2}\,|b|\,a\,m} \right)^{1/(b+2)}, \\
\begin{equation}
\label{Hpl2E}
E= \frac{b+2}{b} \left( \frac{N^2\,(N-1)^{2-b}\,a^2\,b^2\,Q_\phi^{2b}}{16\,m^b} \right)^{1/(b+2)}. \\
\end{equation}
The computation of $\phi$ gives a remarkably simple result
\begin{equation}
\label{Hpl2phi}
\phi=\sqrt{b+2}.
\end{equation}
When $b=2$, the exact result $\phi=2$ is recovered. 

Let us define the ratio $R=\lim_{\nu\gg\lambda}E/\lim_{\lambda\gg\nu}E$. Formulas~(\ref{Hpl2E}) and (\ref{Hpl2phi}) give 
\begin{equation}
\label{Rpl}
R_\textrm{\scriptsize ET}=C_1(b)\left(\frac{\nu}{\lambda}\right)^{2b/(b+2)} \quad\textrm{with}\quad 
C_1(b)=(b+2)^{b/(b+2)}.
\end{equation}
The Hamiltonian~(\ref{Hpl2}) has been studied for $N=2$, $D=3$ and $b>0$ by a Bohr-Sommerfeld quantization (BSQ) approach in \cite{brau00}. From formulas (12) and (14) of this reference, one can get with our notation
\begin{equation}
\label{RBS}
R_\textrm{\scriptsize BSQ}=C_2(b)\,\left(\frac{n_1}{l_1}\right)^{2b/(b+2)} \quad\textrm{with}\quad 
C_2(b)=\frac{2^{2/(b+2)}\,\pi^{2b/(b+2)}\,b^{3b/(b+2)}}{(b+2)\,B(1/b,3/2)^{2b/(b+2)}},
\end{equation}
where $B(x,y)$ denotes the beta function. The dependence on quantum numbers of these two ratios is the same since $\nu = n_1$ and $\lambda = l_1$ asymptotically when $N=2$. The coefficients $C_1(b)$ and $C_2(b)$ are different. But, for $b \in [0,2.5]$, the relative difference $\delta(b)=|C_1(b)-C_2(b)|/(C_1(b)+C_2(b))$ does not exceed 1.6\%, with $\delta(0)=\delta(2)=0$. For larger values of $b$, the situation is less favorable, since $\lim_{b\to\infty}C_1(b)=\infty$ while $\lim_{b\to\infty}C_2(b)=\pi^2$. Nevertheless, the agreement between the two approaches can be considered as quite good, since $\phi$ is computed with the assumption that $\lambda\gg\nu$, not $\nu\gg\lambda$.

We have checked that $\phi=\sqrt{2}$ if $V(r)=a\,\ln(r/b)$. This corresponds to $b=0$ in (\ref{Hpl2phi}), as expected for a logarithmic potential.

\subsection{Power-law 1-body interaction for ultrarelativistic particles}
\label{sec:pl1ur}

In the second case, ultrarelativistic particles ($m=0$) interact via a 1-body power-law potential
\begin{equation}\label{Hpl1}
T(p)=p, \ U(s)=a\,s^b, \ V(r)=0,
\end{equation}
with $a>0$ and $b>0$. The ET gives the following upper bound for the mass $E$ if  $b \le 2$ (if $b > 2$, $E$ has no determined variational character)
%\label{Hpl1r0}
%r_0&=& \left( \frac{N^b\,Q_\phi}{a\, b} \right)^{1/(b+1)}, \\
\begin{equation}
\label{Hpl1E}
E= \frac{b+1}{b} \left( N\,a\,b\,Q_\phi^b \right)^{1/(b+1)}. \\
\end{equation}
Again, a very simple result is obtained
\begin{equation}
\label{Hpl1phi}
\phi=\sqrt{b+1}.
\end{equation}
We have checked that $\phi=1$ if $U(s)=a\,\ln(s/b)$. This corresponds to $b=0$ in (\ref{Hpl1phi}), as expected for a logarithmic potential.

\subsection{Weakly interacting particles}
\label{sec:wib}

Let us now consider nonrelativistic particles interacting via a soft 2-body Gaussian potential
\begin{equation}\label{Hwib}
T(p)=\frac{p^2}{2 m}, \ U(s)=0, \ V(r)=-V_0\,e^{-r^2/R^2}.
\end{equation}
With the definition
\begin{equation}
\label{HwibY}
Y(Z)=-\frac{1}{N^{1/2}\, (N-1)}\frac{Z}{R\, \sqrt{2\,m\, V_0}},
\end{equation}
the resolution of the ET equations gives the following upper bound for the energy $E$ 
\begin{equation}
%\label{Hwibr0}
%r_0&=& N^{1/2}(N-1)^{1/2}\, \sqrt{-W_0(Y)}\, R,\\
\label{HwibE}
E=-\frac{N(N-1)}{2}\, V_0\, Y(Q_\phi)^2\, \frac{1+2\, W_0(Y(Q_\phi))}{W_0(Y(Q_\phi))^2},
\end{equation}
where the multivalued Lambert function $W(z)$ is the inverse function of $z\,e^z$ \cite{corl96}. $W_0(z)$ is the branch defined for $z \ge -1/e$. The computation of $\phi$ gives
\begin{equation}
\label{Hwibphi}
\phi=2\sqrt{1+W_0(Y(\lambda))}.
\end{equation}
When $R\to\infty$, the interaction reduces to a constant plus a harmonic potential, $V(r)\approx -V_0(1-r^2/R^2)$, in the physical range of the eigenstates. It can be checked that $E\to -C_N\,V_0+\sqrt{2\,N\,V_0/m\,R^2}\,Q_\phi$ and $\phi\to 2$, as expected.  

The good quality of the ET results for this kind of systems has been checked in \cite{sema15} for the ground state of bosons, thanks to the accurate numerical results obtained in \cite{horn14,timo12}. With the parameters used in \cite{sema15}, (\ref{Hwibphi}) gives values of $\phi$ varying from 1.85 to 1.95, when $N$ runs from 2 to 20. This is in good agreement with the constant value $\phi=1.82$ fitted in \cite{sema15}.

\subsection{Self-gravitating particles} 
\label{sec:sgb}

The following Hamiltonian,
\begin{equation}\label{Hsgb}
T(p)=\frac{p^2}{2 m}, \ U(s)=0, \ V(r)=-\frac{g}{r},
\end{equation}
is a particular case of (\ref{Hpl2}) with  $b=-1$ and $\phi=1$. For a self-gravitating system, $g=m^2 G$ where $G$ is the gravitational constant. Thanks to the accurate numerical results obtained in \cite{horn14}, it has been shown in \cite{sema15} that the ET results for the ground state of systems composed of up to 8 bosons can be largely improved by a value of $\phi$ around 1. The value $\phi=1$ also allows a good agreement between the ET prediction and a lower bound for the ground state of $N$ bosons in the $D=3$ space \cite{basd90}.

\subsection{Confined particles} 
\label{sec:cb}

Now, we consider particles confined by a harmonic oscillator potential with a pairwise repulsive Coulomb interaction
\begin{equation}\label{Hcb}
T(p)=\frac{p^2}{2 m}, \ U(s)=\frac{1}{2}m\,\omega^2 s^2, \ V(r)=\frac{g}{r}.
\end{equation}
With the definition
\begin{equation}
\label{HcbY}
Y(Z)=\frac{2^{16/3}}{3}\frac{1}{N^{4/3}(N-1)^2} \left(\frac{\omega}{m\, g^2}\right)^{2/3} Z^2,
\end{equation}
the ET method gives the following lower bound
\begin{equation}
%\label{Hcbr0}
%r_0&=& \frac{N^{5/6}(N-1)^{1/2}}{2^{5/6}} \left(\frac{g}{m\, \omega^2}\right)^{1/3} G_{-}(Y),\\
\label{HcbE}
E=\frac{N^{2/3}(N-1)}{2^{5/3}} \left( m\, \omega^2\, g^2\right)^{1/3} \left( G_{-}(Y(Q_\phi))^2 + G_{-}(Y(Q_\phi))^{-1} \right),
\end{equation}
where $G_{\pm}(Y)$ is the only positive root of the quartic equation $4 x^4 \pm 8x - 3Y =0$  with $Y \ge 0$ \cite{silv12} \footnote{$G_{\pm}(Y) = \mp \frac{1}{2} \sqrt{V(Y)} + \frac{1}{2} \sqrt{ 4 (V(Y))^{-1/2}- V(Y)}$ with 
$V(Y)=\left(2 + \sqrt{4 + Y^3} \right)^{1/3} -  Y\left(2 + \sqrt{4 + Y^3} \right)^{-1/3}$.}. 
For the bosonic ground state, the value $3\,\omega/2$ must be added to (\ref{HcbE}) if the confinement is ruled by $\sum_{i=1}^N \bm r_i^2$ instead of $\sum_{i=1}^N (\bm r_i-\bm R)^2$. The computation of $\phi$ gives
\begin{equation}
\label{Hcbphi}
\phi= 2 \sqrt{\frac{2\,G_{-}(Y(\lambda))}{Y(\lambda)}+1}.
\end{equation}
One can check that $E=\omega\, Q_\phi$ with $\phi=2$ when $g=0$, which is the exact solution in this case.  

\begin{figure*}[htb]
\includegraphics[width=0.48\textwidth]{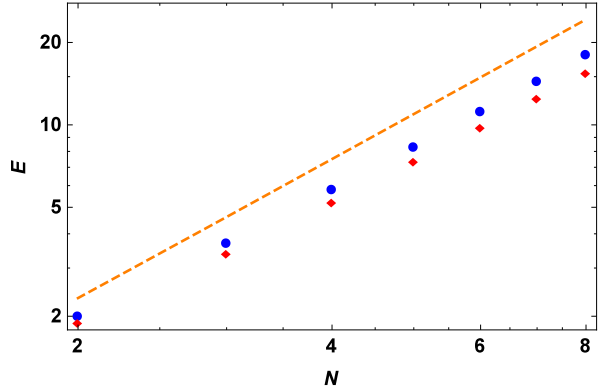} 
\caption{Log-log plot of energies, $E$, of $N$ confined bosons: exact results (circle) \cite{horn14}, ET results for $\phi=2$ (diamond), ET results for $\phi$ given by (\ref{Hcbphi}) (dashed line).}
\label{fig:cbE} 
\end{figure*}

The comparison with the accurate numerical results obtained in \cite{horn14} for the ground state of systems composed of up to 8 bosons with the parameter $m=1$, $\omega=0.5$ and $g=1$ shows that the ET gives good lower bounds. Better results for the ET can be obtained by increasing, with the number of particles, the value of $\phi$ above 2. This is achieved by formula~(\ref{Hcbphi}), but too much to give an improvement of the results. This is illustrated on Fig.~\ref{fig:cbE}. 

\subsection{Large-$N$ baryons}
\label{sec:lnb}

In a particular large-$N$ limit of QCD ($N$ is the number of colors), light baryons are composed of $N$ ultrarelativistic quarks $u$ or $d$ ($m=0$) in the fundamental color representation. In this case, the corresponding Hamiltonian is \cite{buis12}
\begin{equation}\label{Hlnb}
T(p)=p, \ U(s)=k\, s, \ V(r)=-\frac{g}{r}.
\end{equation}
Quarks are fermions but they can be treated as bosons since the color part of the wave-function is completely antisymmetrical. With the ET, the following upper bound is obtained for the mass $E$ of the baryons
\begin{equation}
%\label{Hlnbr0}
%r_0&=& \frac{1}{\sqrt{\lambda}}\,\sqrt{ N\, Q_\phi - \left(\frac{N(N-1)}{2}\right)^{3/2} g }, \\
\label{HlnbE}
E= \sqrt{4\,k}\,\sqrt{ N\, Q_\phi - \left(\frac{N(N-1)}{2}\right)^{3/2} g }.
\end{equation}
The computation of $\phi$ gives
\begin{equation}
\label{Hlnbphi}
\phi=\sqrt{2-\frac{\sqrt{N\,(N-1)^3}\,g}{\sqrt{2}\,\lambda}}.
\end{equation}
These results with $g=0$ are in agreement with those of Sec.~\ref{sec:pl1ur} with $b=1$.

In order to test the relevance of $Q_\phi$ for the ground state and some excited states, we consider the 3-body system studied in \cite{silv10}. For the genuine ET ($\phi=2$), the mean relative error $\Delta$ computed on the 16 eigenmasses listed in Table~\ref{tab:1} is 15.1\%. With the value $\phi$ predicted by (\ref{Hlnbphi}), $\Delta$ reduces to 4.7\%. More precisely, the mean error on the 4 states of Table~\ref{tab:1} with $l_1+l_2=0$ is 9.1\%, while it is only 3.2\% on the 12 remaining states. This could be interpreted as a trace of the condition of validity of the DOS method ($\lambda \gg 1$). 

Lets us remark that better global agreements can be obtained with constant values of $\phi$. With $\phi=1.35$, the ET yields the exact value for the ground states and $\Delta$ drops to 3.1\%. Finally, with $\phi=1.23$, the minimal value of 2.4\% is reached for $\Delta$. Obviously, it is necessary to known some exact solutions of the system to compute these values of $\phi$. From Table~\ref{tab:1}, it is also clear that the variational character of the ET solutions is lost when $\phi\ne 2$.

\begin{table}[htbp]
\caption{Some eigenmasses in GeV of the Hamiltonian~(\ref{Hlnb}) for $D=3$, $N=3$, $k=0.2$~GeV and $g=\frac{2}{3}\alpha_S$ with $\alpha_S=0.4$  \cite{silv10}. Accurate results obtained from an expansion in a harmonic oscillator basis \cite{silv96} are given in the column ``Exact". Results computed with the ET for four values of $\phi$ are listed in the four last columns, with the associated mean relative errors $\Delta$. The sums $n_1+n_2$ and $l_1+l_2$ are the quantum numbers of the main component of the corresponding genuine eigenstate in the harmonic oscillator basis. These numbers are used for the computation of $Q_\phi$.}
\begin{center}
\label{tab:1}
\begin{tabular}{lllllll}
\hline\noalign{\smallskip}
 & & & \multicolumn{4}{c}{ET} \\
$n_1+n_2$ & $l_1+l_2$ & Exact & $\phi=2$ & (\ref{Hlnbphi}) & $\phi=1.35$ & $\phi=1.23$ \\
\noalign{\smallskip}\hline\noalign{\smallskip}
0 & 0 & 2.128 & 2.468 & 1.945 & 2.128 & 2.060 \\
0 & 1 & 2.606 & 2.914 & 2.582 & 2.633 & 2.578 \\
1 & 0 & 2.739 & 3.300 & 2.504 & 2.788 & 2.682 \\
0 & 2 & 2.959 & 3.300 & 3.035 & 3.055 & 3.007 \\
1 & 1 & 3.125 & 3.646 & 3.106 & 3.189 & 3.098 \\
0 & 3 & 3.299 & 3.646 & 3.418 & 3.425 & 3.383 \\
2 & 0 & 3.260 & 3.961 & 2.960 & 3.318 & 3.186 \\
1 & 2 & 3.422 & 3.961 & 3.512 & 3.546 & 3.463 \\
0 & 4 & 3.581 & 3.961 & 3.758 & 3.759 & 3.721 \\
2 & 1 & 3.584 & 4.253 & 3.553 & 3.662 & 3.542 \\
1 & 3 & 3.716 & 4.253 & 3.857 & 3.869 & 3.794 \\
0 & 5 & 3.861 & 4.253 & 4.068 & 4.066 & 4.030 \\
3 & 0 & 3.721 & 4.527 & 3.354 & 3.775 & 3.619 \\
2 & 2 & 3.838 & 4.527 & 3.932 & 3.976 & 3.866 \\
1 & 4 & 3.966 & 4.527 & 4.166 & 4.168 & 4.098 \\
0 & 6 & 4.103 & 4.527 & 4.356 & 4.351 & 4.318 \\
\noalign{\smallskip}\hline\noalign{\smallskip}
$\Delta$ & & & 15.1\% & 4.7\% & 3.1\% & 2.4\% \\
\noalign{\smallskip}\hline
\end{tabular}
\end{center}
\end{table}

\section{Conclusion}
\label{sec:conclu}

The envelope theory is a powerful method to compute eigenvalues for quite general $N$-body systems with identical particles in $D$ dimensions \cite{sema13a}. The method is easy to implement and can produce analytical upper or lower bounds in favorable situations, as those presented in this paper. Its purpose is not to compete with accurate methods such as the explicitly correlated Gaussian basis \cite{suzu98} or the harmonic oscillator basis \cite{silv96}, but to produce without great pain reliable estimations for the energy. 

It has been shown in \cite{sema15} that the eigenvalues computed with the envelope theory can be quite close to the exact results, and that they can be improved by the introduction of a single parameter $\phi$ in the characteristic global quantum number of the method. This procedure has been inspired from \cite{loba09}, where a universal effective quantum number for centrally symmetric 2-body systems is proposed. The price to pay is that the variational character of the approximation cannot then be guaranteed. This parameter $\phi$ can be determined, for instance, by the knowledge of an exact solution for $N=2$ or by a fit on numerical results. 

In this paper, a method to compute the value of $\phi$, independently of the knowledge of analytical or numerical previous calculations, is proposed by using, in the envelope theory, information coming from a $N$-body generalization of the dominantly orbital state method \cite{olss97,sema13b}. A procedure to compute $\phi$ is given in the general case. Though, the complete formula is quite complicated, it can reduce to very simple forms, as for the cases presented in this paper. 

From the way $\phi$ is computed, it is expected that the improvement of the eigenvalues is good only when the global angular momentum is large and the global radial excitation is small. But the situation is not so clear, since even ground states can be improved. With a value $\phi\ne 2$, the variational character of the approximation cannot be guaranteed. It is then not possible to locate with certainty the approximation with respect to the exact energy. Nevertheless, the artificial strong degeneracy inherent to the envelope theory is broken, and relevant insights about the relations between orbital and radial excitations can be obtained. 

Since $N$-body problems are always difficult and heavy to solve accurately, the envelope theory, supplemented by the computation of $\phi$, can be used as a guide for the study of these complicated systems. It could be interesting to test this method with other systems in order to better specify the domain of validity of this procedure. Accurate numerical results for various systems, with different values of $N$ and $D$, are welcome.

\end{document}